\documentclass[]{spie}

\usepackage[utf8]{inputenc}

\usepackage{tikz}
\usepackage{pgfplots}
\pgfplotsset{compat=1.5}

\usepackage{amsmath,amsfonts,amssymb}

\usepackage{graphicx}
\usepackage[colorlinks=true, allcolors=blue]{hyperref}

\usepackage{txfonts}
\usepackage{textcomp}
\usepackage{eurosym}

\newcommand{\micron}{\mbox{$\mu$m}}

\title{FRIDA: diffraction-limited imaging and integral-field spectroscopy for the GTC}

\author[a]{Alan~M.~Watson}
\author[b]{José~A.~Acosta-Pulido}
\author[a]{Luis~C.~Álvarez-Núñez}
\author[c]{Vicente~Bringas-Rico}
\author[d]{Nicolás~Cardiel}
\author[a]{Salvador~Cuevas}
\author[a]{Oscar~Chapa}
\author[b]{José~Javier~Díaz~García}
\author[e]{Stephen~S.~Eikenberry}
\author[a]{Carlos~Espejo}
\author[a]{Rubén~A.~Flores-Meza}
\author[a]{Jorge~Fuentes-Fernández}
\author[d]{Jesús~Gallego}
\author[a]{José~Leonardo~Garcés~Medina}
\author[b]{Francisco~Garzón~López}
\author[f]{Peter~Hammersley}
\author[a]{Carolina~Keiman}
\author[a]{Gerardo~Lara}
\author[g]{José~Alberto~López}
\author[b]{Pablo~L.~López}
\author[c]{Diana~Lucero}
\author[b]{Heidy~Moreno~Arce}
\author[d]{Sergio~Pascual~Ramirez}
\author[b]{Jesús~Patrón~Recio}
\author[b]{Almudena~Prieto}
\author[c]{Alberto~José~Rodríguez}
\author[b]{José~Marco~de~la~Rosa}
\author[a]{Beatriz~Sánchez}
\author[c]{Jorge~A.~Uribe}
\author[c]{Francisco~Váldez~Berriozabal}

\affil[a]{Instituto de Astronomía, Universidad Nacional Autónoma de México, Apartado Postal 70-264, 04510~México, Mexico}
\affil[b]{Instituo de Astrofísica de Canarias, c/Vía Láctea s/n, E-38200, La Laguna, Tenerife, Spain}
\affil[c]{Centro de Ingeniería y Desarrollo Industrial, Avenida Playa Pie de la Cuesta 702, 76130 Querétaro, Mexico}
\affil[d]{Departamento de Física y Astronomía, Universidad Complutense de Madrid, Avenida de Séneca 2, Ciudad Universitaria, 28040 Madrid, Spain}
\affil[e]{Astronomy Department, University of Florida, 211 Bryant Space Center, PO Box 112055, Gainsville, FL 32611-2055, USA}
\affil[f]{European Southern Observatory, Karl Schwarzschild Straße 2, 85748 Garching, München, Germany}
\affil[g]{Instituto de Astronomía, Universidad Nacional Autónoma de México, Apartado Postal 106, 22860~Ensenada, Baja California, Mexico}

\authorinfo{Further author information: Send correspondence to A.M.W. E-mail: alan@astro.unam.mx}

\begin{document} 
\maketitle

\begin{abstract}
FRIDA is a diffraction-limited imager and integral-field spectrometer that is being built for the adaptive-optics focus of the Gran Telescopio Canarias. In imaging mode FRIDA will provide scales of 0.010, 0.020 and 0.040 arcsec/pixel and in IFS mode spectral resolutions of 1500, 4000 and 30,000. FRIDA is starting systems integration and is scheduled to complete fully integrated system tests at the laboratory by the end of 2017 and to be delivered to GTC shortly thereafter. In this contribution we present a summary of its design, fabrication, current status and potential scientific applications.
\end{abstract}


\section{INTRODUCTION}

FRIDA (InFRared Imager and Dissector for the Adaptive optics system of GTC) is a diffraction-limited instrument that will operate in the 0.9--2.5 {\micron} wavelength range. FRIDA has been designed and is being built as a collaborative project between institutions from Mexico, Spain and the USA.  FRIDA will be located at the Nasmyth B focus of GTC, behind the GTCAO adaptive optics system. FRIDA has completed its critical design review and an intensive phase of prototypes testing and design optimization. FRIDA is now starting systems integration and is scheduled to complete fully integrated system tests and verification at the laboratory by the end of 2017 and be delivered to GTC shortly after. The availability of FRIDA on GTC will then depend on the delivery of GTCAO (scheduled for late 2018) and commissioning by GTC staff.

FRIDA will offer broad and narrow band imaging with selectable spatial scales of 0.010 and 0.020 arcsec per pixel corresponding to fields of view of 20.48 x 20.48 arcsec and 40.96 x 40.96 arcsec, respectively. The finer scale will provide adequate sampling for the nearly diffraction-limited core in $J$ and $H$ bands and the medium scale will provide adequate sampling in K band. A third camera with the coarser 0.040 arcsec per pixel scale will be available to aid acquisition in the IFS mode and for poor AO correction situations.

\section{OBSERVING CAPABILITIES}

FRIDA will provide imaging and integral-field spectroscopy on the same HAWAII-II detector (although see \S\ref{section:detector} on the upgrade path to a HAWAII-2RG).

\subsection{Imaging}

FRIDA will have three cameras for imaging: the fine camera will have 0.010 arcsec pixels and a field of 20 arcsec; the medium camera will have 0.020 arcsec pixels and a field of 40 arcsec; and the coarse camera will have 0.040 arcsec pixels but will only use the central portion of the detector and will be limited again to a field of 40 arcsec. The fine and medium cameras will provide Nyquist sampling of diffraction-limited images above 1.0 and 2.0 {\micron} respectively. Their nominal fields are sufficiently large that the effective field will likely be limited by anisoplanetism for the singly-conjugate GTCAO adaptive optics system. Imaging with the coarse camera is mainly intended to be used for acquisition for spectroscopy with the same camera.

FRIDA has two filters each with space for 18 filters. We plan to equip it with wide-band $ZJHK$, $J_\mathrm{s}$, and $K_\mathrm{s}$ filters, a selection of narrow-band and medium-band filters targeting interstellar emission lines and stellar absorption features, neutral density filters, and order-sorting filters for spectroscopy.

\subsection{Integral-Field Spectroscopy}

FRIDA performs spectroscopy with an image slicer whose geometry is 64 pixels along each of 30 slits. The slits are two pixels wide, both to give adequate spectral sampling and to avoid the need for excessively oversized optics in the spectrograph to mitigate diffraction. Each “spaxel” or unit of spatial resolution is this a rectangle with a 2:1 aspect ratio with the shorter edge being parallel to the slit.

As in imaging mode, three cameras give different scales. The fine camera gives spaxels of $0.010 \times 0.020$ arcsec and a field of $0.60 \times 0.64$ arcsec, the medium camera gives spaxels of $0.020 \times 0.040$ arcsec and a field of $1.20 \times 1.28$ arcsec, and the coarse camera gives spaxels of $0.040 \times 0.080$ arcsec and a field of $2.40 \times 2.56$ arcsec. The fine and medium cameras will provide Nyquist sampling of diffraction-limited images along the slit above 1.0 and 2.0 {\micron} respectively. The fine camera gives Nyquist sampling of diffraction-limited images perpendicular to the slit above 2.0 {\micron}. The coarse camera is for applications that wish to trade spatial sampling for sensitivity or field.

Low, medium, and high-resolution gratings are available. The low-resolution grating gives $R \approx 1500$ in the combined $ZJ$ or $HK$ windows. The medium-resolution gratings give $R\approx 4000$ in each of the $Z$, $J$, $H$, and $K$ windows. The high-resolution grating gives $R \approx 30,000$ in $\lambda/30$ regions of the $H$ and $K$ windows.

\section{DESIGN}

\subsection{Conventional Optics}

\begin{figure}
\begin{center}
\includegraphics[width=0.8\linewidth]{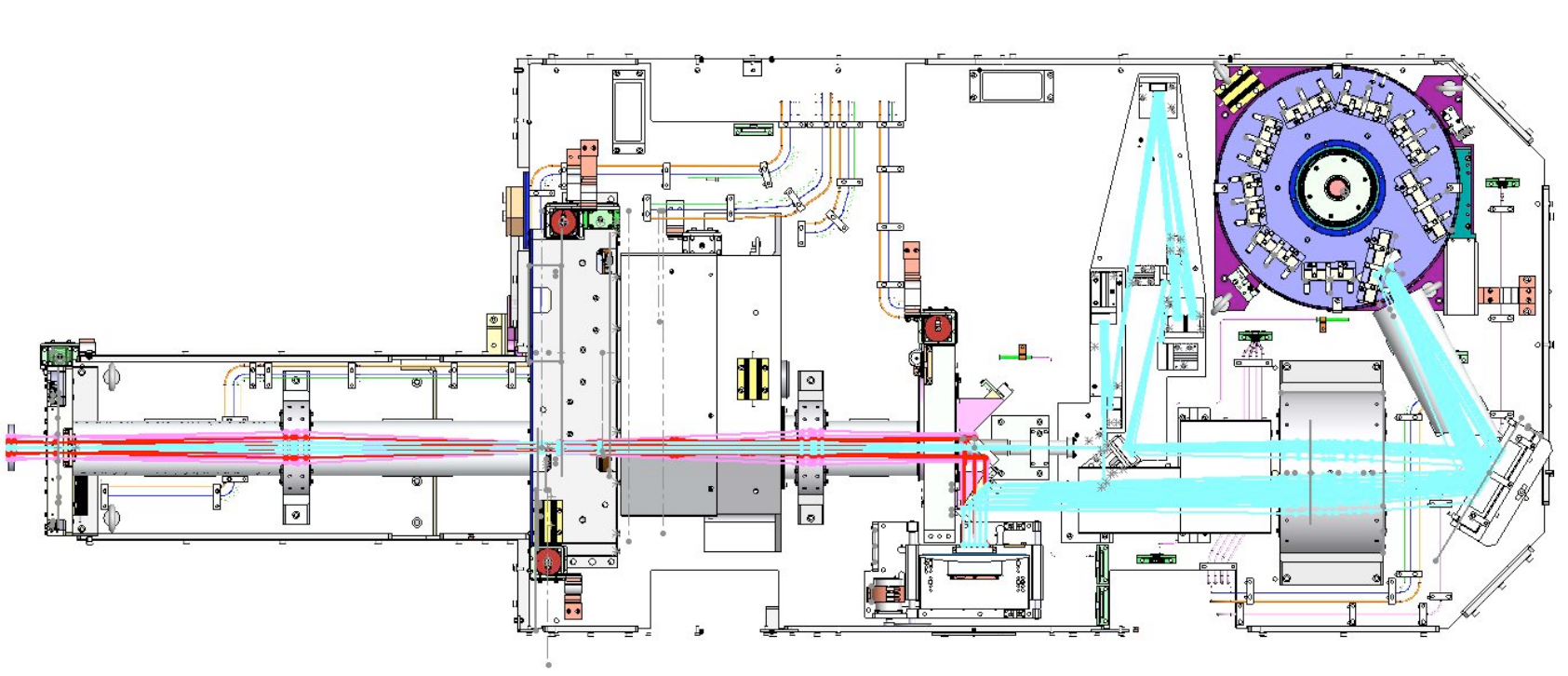}
\end{center}
\caption{The optical paths in FRIDA. The beams enter from the left from GTC and the GTCAO adaptive optics system. The red beam shows the path in imaging mode, and the cyan beam shows the path in integral-field spectroscopy mode. The beams follow identical paths through the fore-optics. The mechanism at the point that the red and cyan beams diverge is the mode selection mechanism. The detector is in the lower middle. In imaging mode, the fore-optics reimages the input focal plane onto the detector. In integral-field spectroscopy mode, the fore-optics images the input focal plane onto the image slicer, the integral-field unit the reimages the slices into a pseudo slit, and the spectrograph reimages the pseudoslit onto the detector.}
\label{figure:optical-design}
\end{figure}

Figure~\ref{figure:optical-design} illustrates the optical design of FRIDA. FRIDA receives a corrected and derotated focal plane image from GTC and the GTCAO adaptive optics system. This focal plane is then reimaged once in imaging mode or four times in integral field spectroscopy mode.

The FRIDA optical design uses common fore-optics for both imaging and spectroscopy. These consist of a fixed collimator and selectable fine, medium, and coarse science cameras and an engineering pupil viewing camera. The collimator forms an image of the pupil and a mechanism can select different pupil masks (inscribed, circumscribed, and intermediate). Focal plane masks can also be placed in the input focal plane, to help control scattered light and for calibration. The focal plane and pupil plane mechanisms provide a simple upgrade path to coronography.

After the fore-optics, a mechanism places flat mirrors into the beam to switch between imaging and integral-field spectroscopy modes. In imaging mode, the focal plane formed by the fore-optics is focused onto the detector. In integral-field spectroscopy mode, it is reimaged onto the image slicer, then the image slicer is reimaged into a single pseudoslit, and finally the pseudoslit is reimaged by the double-pass spectrograph optics onto the detector. 

The slicer has 30 slices each of 64 pixels, and each slice is imaged onto the detector with wavelengths being approximately continuous from one slice to another. Four pixels are left between slices to reduce crosstalk. The spectrograph grating turret holds eight gratings working at a near Littrow configuration and two mirrors (for field acquisition through the slicer).

All the refractive components in the collimator, cameras, and the spectrograph are based on airspace doublets using two cryogenic matching materials, namely, CaF$_2$ and Ohara S-FTM16. The spectrograph also employs an Infrasil-301 lens.

In imaging mode, the Strehl ratio degradation is expected to be better than 0.90 from 1.1 to 2.4 {\micron} and better than 0.8 from 0.9 to 2.5 {\micron}. In integral-field spectroscopy mode, it is expected to be better than 0.8 from 1.1 to 2.4 {\micron}. The throughput is expected to be better than 60\% in imaging mode (excluding filters and the detector) and better than 25\% in integral-field spectroscopy mode (excluding filters, gratings, and the detector).

\subsection{Integral Field Unit}

\begin{figure}
\begin{center}
\includegraphics[width=0.8\linewidth]{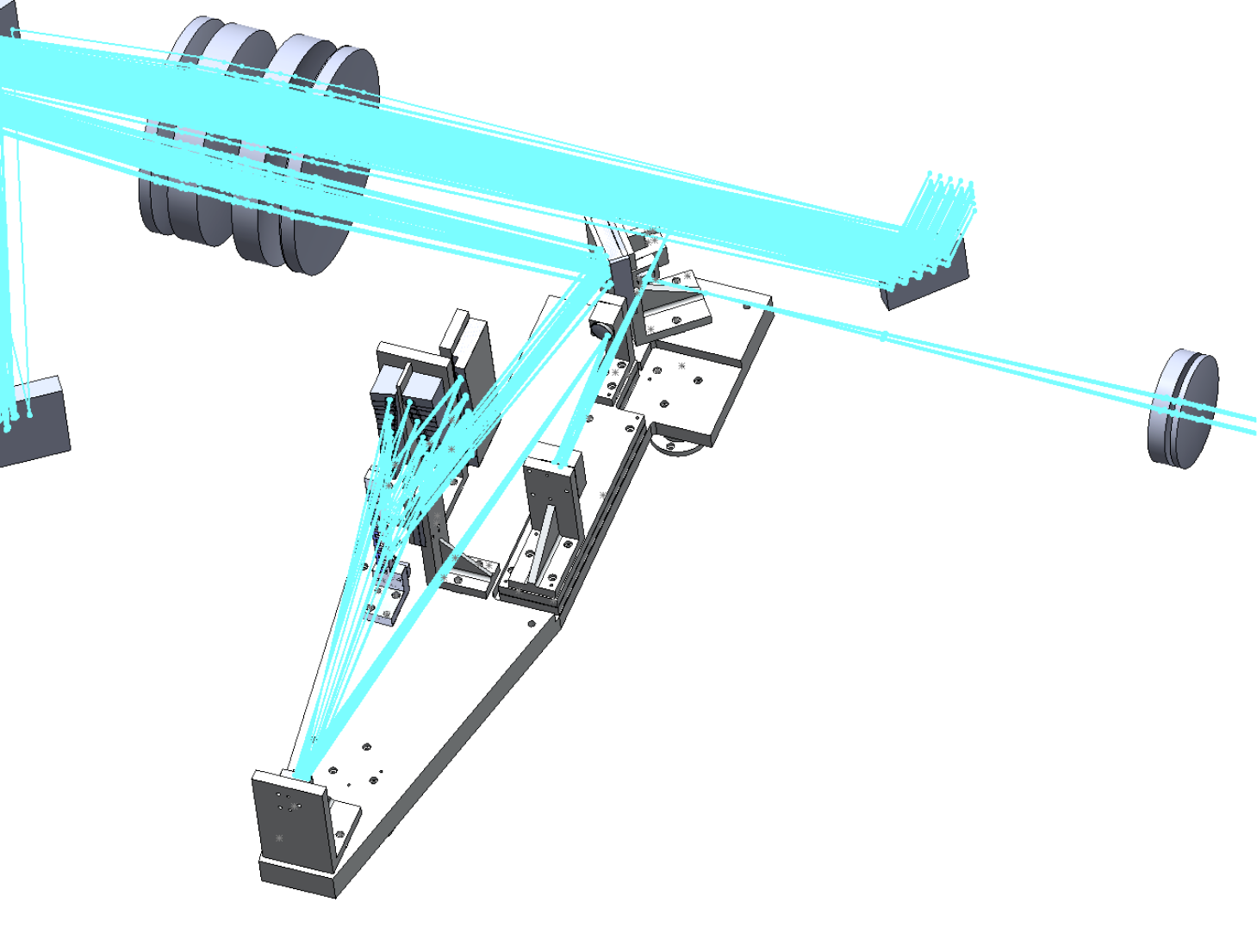}
\end{center}
\caption{A rendered image of the IFU and the spectrographs optics. The beam is shown in cyan and enters from the right. The image slicer is located in the lower part of the image, the spectrograph in the upper left of the image, and the detector in the upper part of the image.}
\label{figure:ifu-design}
\end{figure}

The design of the FRIDA integral-field unit (IFU) was performed by the UNAM and the University of Florida and is based on the University of Florida’s FISICA IFU. The IFU mirrors are diamond-turned into RSA 6061 aluminum coated with a nickel alloy to reduce scattering. 

Figure~\ref{figure:ifu-design} shows in detail the optical layout for the integral field unit and the spectrograph. The beam enters the IFU through the IFU focal point, a Schwarzschild relay performs an amplification before the beam reaches the slicer mirrors block. The sliced image is sent to two sets of pupil mirrors blocks that redirect it to the field mirrors block that arranges the sliced up image into the output pseudo slit image. Two off-axis parabolic mirrors are used to de-magnify and extract the pseudoslit image from the IFU. The rest of the figure shows the beam’s path through the spectrograph, the diffraction grating and back towards the detector. 

\subsection{Mechanical Design}

\begin{figure}
\begin{center}
\includegraphics[width=0.8\linewidth]{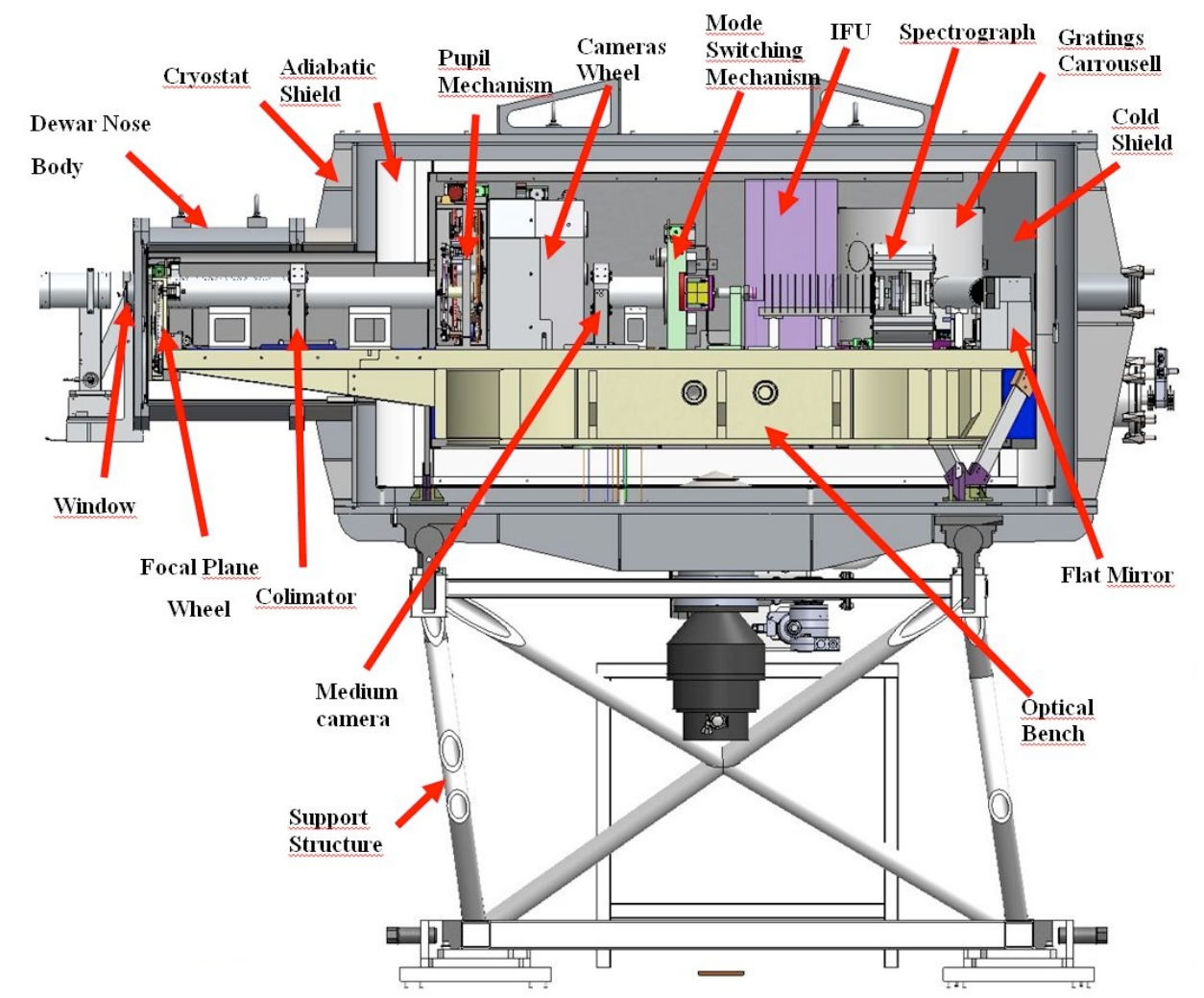}
\end{center}
\caption{A rendered lateral section of FRIDA showing the main optical and mechanical components.}
\label{figure:lateral-section}
\end{figure}

Figure~\ref{figure:lateral-section} shows a lateral section of FRIDA in its cryostat and its main mechanical and optical components. To avoid an interference with the rather inconvenient GTCAO mechanical envelope, the cryostat is extended towards the input focal plane by a “nose”.

\section{CURRENT STATUS AND PLANS}

\subsection{Management}

After leading the project for 10 years, José Alberto López stood down as Principal Investigator at the end of 2014. Alan Watson subsequently took on this rôle.

\subsection{Conventional Optics}

\begin{figure}
\begin{center}
\includegraphics[height=0.25\linewidth]{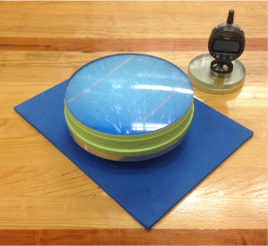}
\includegraphics[height=0.25\linewidth]{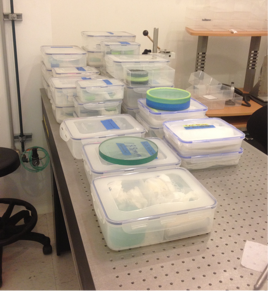}
\end{center}
\caption{A polished lens (left) and the entire set of polished lenses (right).}
\label{figure:optics-photos}
\end{figure}

The manufacturing of the conventional optical elements was carried out at the Instituto de Astronomía of the UNAM using a Strasbaugh generating machine, three Strasbaugh polishing machines, and a Satisloh rounding machine. Surface quality, curvature and rugosity evaluation have been performed with a Zygo 6-inch XP  interferometer. All 25 lenses, mirrors, and windows have now been polished, characterized for figure and microroughness, rounded and bezeled, and are awaiting final cosmetic characterization. We expect to send them for coating in summer 2016.

\subsection{Integral Field Unit}

\begin{figure}
\begin{center}
\includegraphics[height=0.4\linewidth]{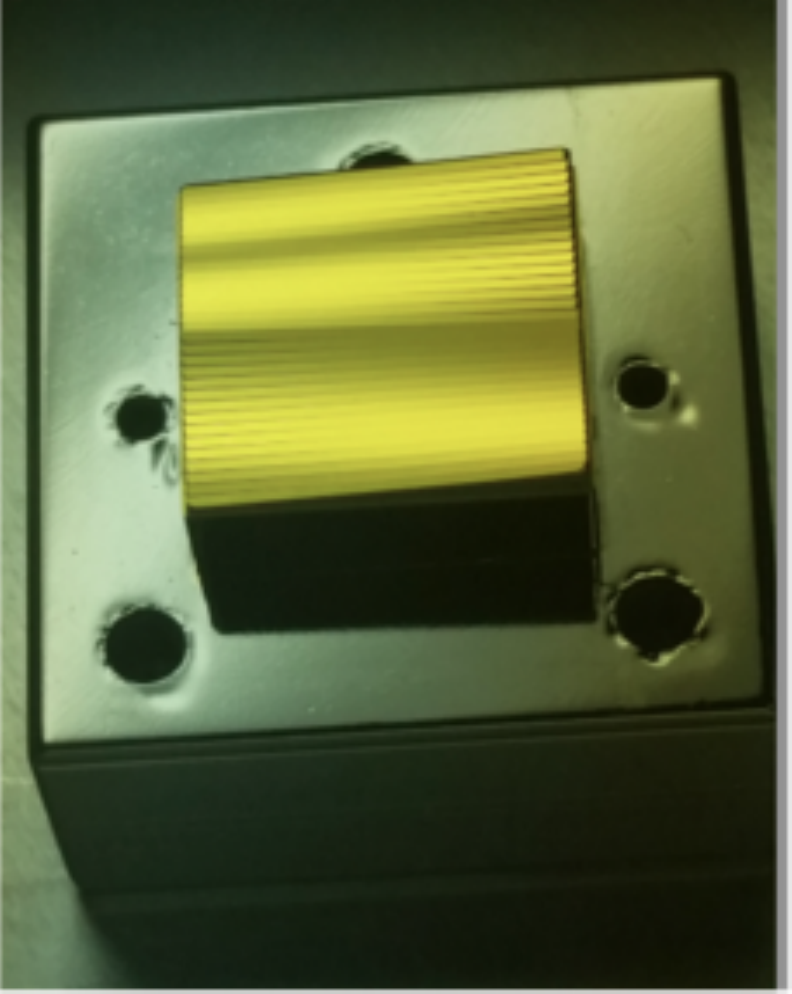}
\includegraphics[height=0.4\linewidth]{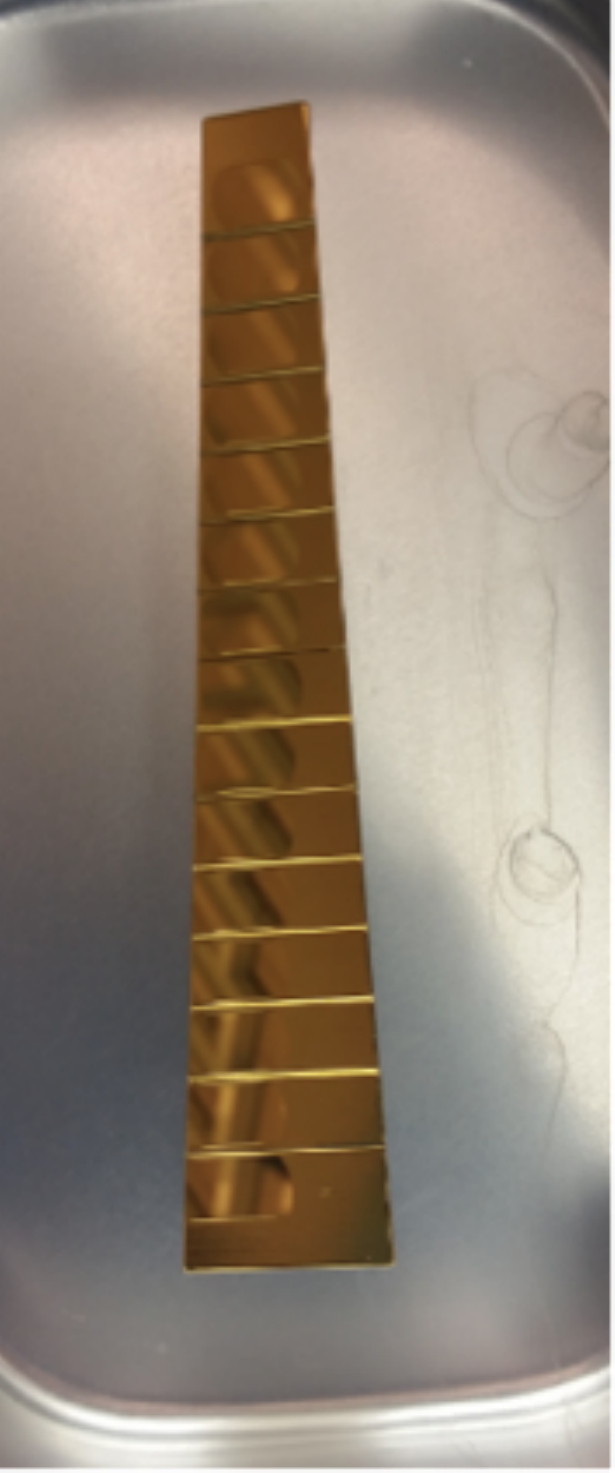}
\includegraphics[height=0.4\linewidth]{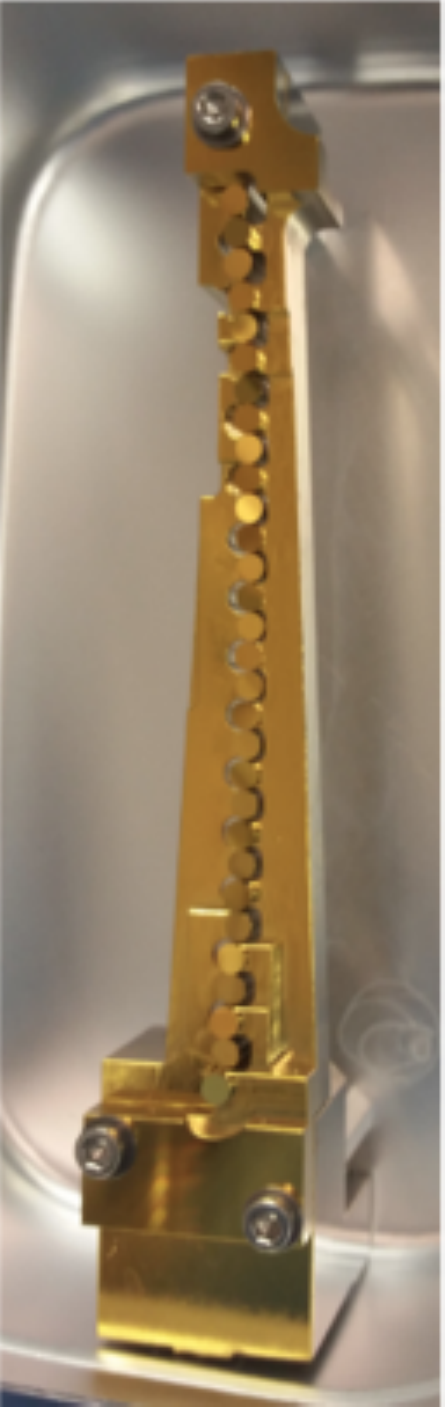}
\end{center}
\caption{Delivered IFU optics. The slicer-mirror array (left), pupil-mirror array (middle), and field-mirror array (right).}
\label{figure:ifu-photos}
\end{figure}

The optics of the integral field unit have now been delivered by Corning NetOptix and Durham Precision Optics. They are currently at the University of Florida awaiting integration on their mechanical bench and subsystem testing. Figure~\ref{figure:ifu-photos} shows the slicer-mirror array, the pupil-mirror arrays, and the field-mirror arrays.

\subsection{Mechanics}

\begin{figure}
\begin{center}
\includegraphics[height=0.25\linewidth]{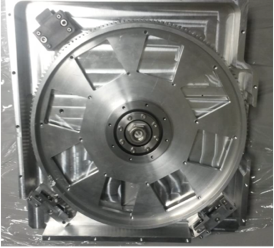}
\includegraphics[height=0.25\linewidth]{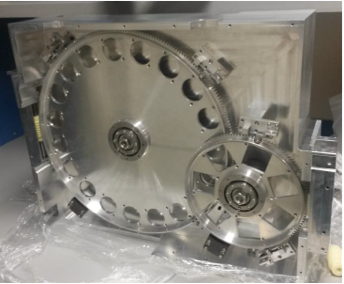}
\includegraphics[height=0.25\linewidth]{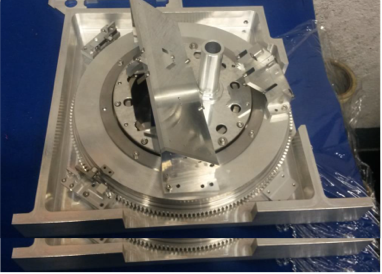}
\end{center}
\caption{Delivered mechanisms. The focal-plane mask wheel (left), pupil-plane mask  and filter wheels (middle), and mode-switching mechanism (right).}
\label{figure:mechanisms-photos}
\end{figure}

\begin{figure}
\begin{center}
\includegraphics[height=0.25\linewidth]{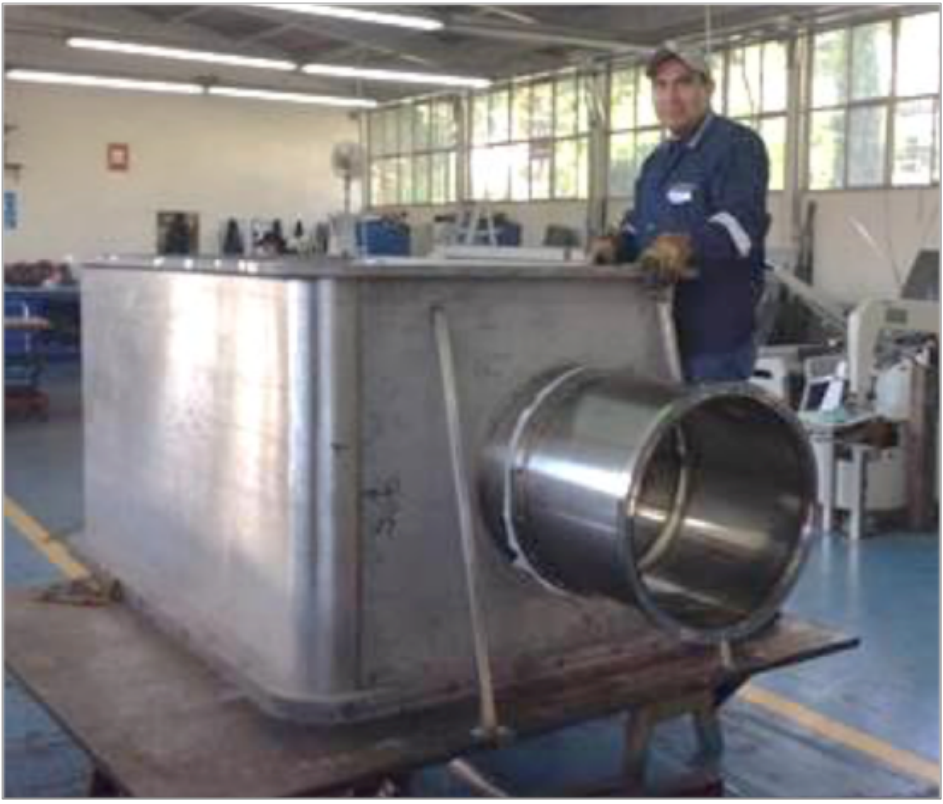}
\includegraphics[height=0.25\linewidth]{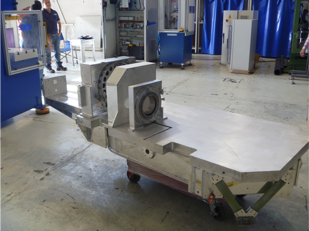}
\includegraphics[height=0.25\linewidth]{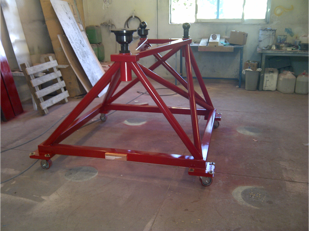}
\end{center}
\caption{The cryostat (left), optical bench (middle), and support structure (left) being finished at CIDESI..}
\label{figure:cryostat-photos}
\end{figure}

The optomechanical design relies on the doublets maintaining their alignment over thermal cycles. We initially had some problems with this due to an insufficiently smooth finish on the metal surface that rests against the lenses. We identified and resolved this problem in 2015 and can now maintain alignment to within 20 {\micron} over thermal cycles. This is well within our tolerance of 85 {\micron}.

The cryostat and optical bench being finished at CIDESI. Many mechanisms have now been delivered and the remainder are expected before the end of 2016. Figure~\ref{figure:mechanisms-photos} and \ref{figure:cryostat-photos} shows some of the components.

\subsection{Control}

The control system is divided between the Instituto de Astrofísica de Canarias (IAC) and the Instituto de Astronomía of the UNAM. 

The IAC is in charge of the hardware and software for the high level control of mechanisms, the data acquisition system, and the detector control. All these tasks are in an advanced stage. The high-level control system is largely based on that of the EMIR instrument\cite{Garzon-2016}. The IAC have made progress in defining the “panel” (GUI screens) for FRIDA. The detector control system is a clone of that for EMIR (although see \S\ref{section:detector} on the upgrade path to a HAWAII-2RG). The IAC has manufactured the thermal connections for the detector fan-out, installed it in a dummy fan-out, and sent this to UF for tests with the focus mechanism. The IAC is also making progress on exposure-time calculator.

The Instituto de Astronomía UNAM is developing the low level, embedded control of mechanisms, and the housekeeping. The most critical mechanisms are focal plane assembly and grating carousel. We have achieved closed-loop control of the carousel (using IAC-compatible interfaces) at ambient temperature and are now  testing at cryogenic temperatures. The UNAM and CIDESI have finalized  the design of the cabling routes and attachment points. The UNAM is currently manufacturing the electronics cabinets.

\subsection{Data Factory Pipeline}

The FRIDA instrument team is  only formally required to provide a “quick look” data pipeline for use at GTC. However, we are aware that the success of FRIDA will depend on having adequate tools for reduction, calibration, and extraction.
Therefore, the UCM group is leading effort to make such tools available. They are producing a package called “pyFRIDA” which will share code and concepts with the similar packages for the EMIR\cite{Garzon-2016} and MEGARA\cite{Gil-de-Paz-2016} instruments. The package is largely designed and partially implemented. Further implementation is on hold pending completion of the EMIR package.

\subsection{Plans}

We plan to being system integration at the UNAM in fall of 2016, system testing in early 2017, laboratory acceptance testing in fall of 2017, and delivery to GTC at the end of 2017. First light will then depend on the availability of the Nasmyth B focal station currently occupied by the OSIRIS instrument, the delivery and commissioning of GTCAO in 2018, and the availability of GTC staff for commissioning. The earliest likely date for first light for FRIDA is early in 2019.

\section{FRIDA and GTCAO}

FRIDA will operate with the GTCAO adaptive optics system of GTC. The development of GTCAO had been under the direct responsibility of GTC. However, due to lack of resources at GTC, this responsibility was transferred to the IAC in 2014. The FRIDA team has no formal involvement in the design and development of GTCAO. Current plans call for GTCAO to be installed at GTC in 2018.

\begin{figure}
\begin{center}
\includegraphics[height=0.25\linewidth]{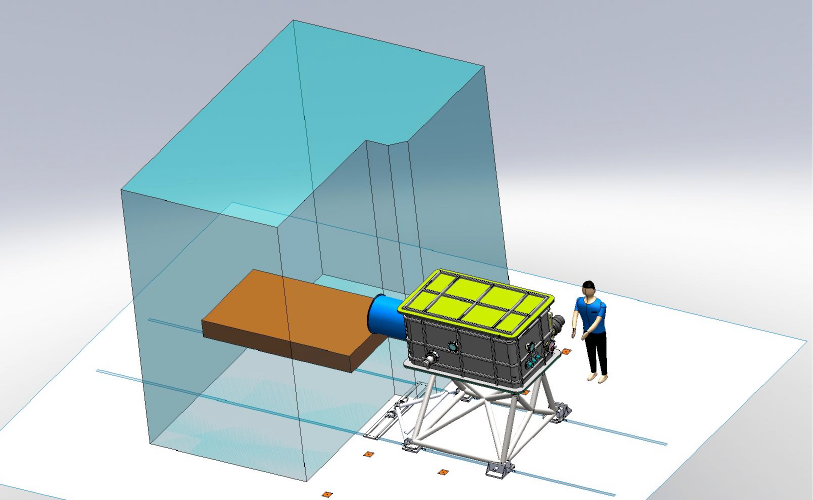}
\includegraphics[height=0.25\linewidth]{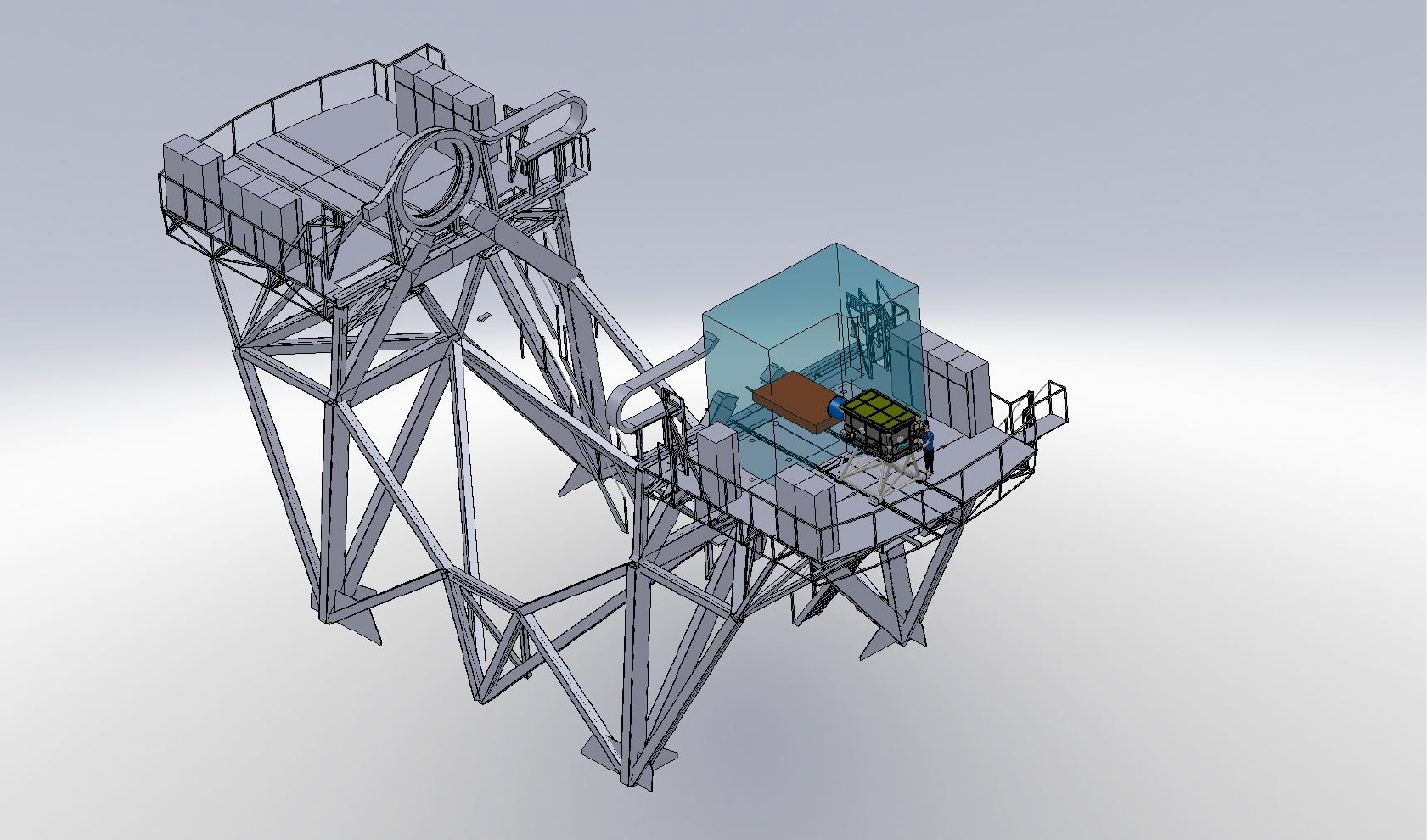}
\end{center}
\caption{Rendered images of FRIDA and GTCAO on the Nasmyth B platform of GTC.}
\label{figure:nasmyth}
\end{figure}

GTCAO will initially operate in the natural guide star mode and provisions are being taken to upgrade it to a laser guide system operation shortly after commissioning of the NGS mode. GTCAO will be a mono-conjugate AO system with a Shack-Hartmann wave front sensor. The deformable mirror has $21 \times 21$ actuators. GTCAO will be a fixed system at the Nasmyth platform with field de-rotator. GTCAO has an extensive envelope in the Nasmyth platform that the FRIDA design must comply with, particularly at the interface between these instruments. Figure~\ref{figure:nasmyth} shows rendered images of FRIDA and the GTCAO envelope at the Nasmyth platform.

\section{DETECTOR}
\label{section:detector}

When FRIDA was conceived, GTC decided that it would use a HAWAII-II detector and a duplicate of the detector control system being developed by the IAC for EMIR. GTC therefore acquired two science-grade HAWAII-II detectors.

It was later discovered that the HAWAII-II detector that was originally intended for EMIR is from a batch that has shown an alarming propensity for catastrophic failures during thermal cycling. Therefore, EMIR was assigned the other detector, the one originally intended for FRIDA, which was from a different batch and was understood to be more robust. Of course, all parties recognized at this action did not solve the problem, but simply transferred the risk of a catastrophic failures from EMIR to FRIDA. However, since FRIDA was at an earlier stage in its development, there was more time to look for a definitive solution. 

An obvious solution would be to acquire a new HAWAII-II device from Teledyne. However, the HAWAII-II was no longer being manufactured, having been superseded by the HAWAII-2RG. At the start of 2013, GTC asked the FRIDA team to evaluate the possibility of changing to an HAWAII-2RG detector. At the time, we stated that a HAWAII-2RG could be installed without any impact on first light at a cost of about 400,000€ and that this was our preferred option. Upon receiving our report and recommendation, GTC decided that FRIDA should stay with a HAWAII-II detector, to maintain commonality with EMIR and to avoid additional costs. In mitigation, GTC undertook to look for a replacement HAWAII-II detector from a batch with a lower risk of failure. GTC secured the loan of what we understand to be the last two HAWAII-II devices available to Teledyne in spring 2015. These were evaluated in the IAC in summer 2015. One was found to be inoperable and the other found to be not a science-grade device.

In November 2015 the FRIDA science team and the GTC Steering Committee agreed that proceeding with the existing HAWAII-II was an unacceptable risk to the huge investment in FRIDA and GTCAO. An effort to look for funds for a HAWAII-2RG is being lead by Almudena Prieto of the IAC under a proposal that shares the cost of the new detector 80\%:20\% between Spain and Mexico. At this late stage, changing to a HAWAII-2RG will almost certainly impact the schedule for delivery, but perhaps not first light, since there is about a year of slack between the delivery of FRIDA and projected first light with GTCAO.

\section{SCIENCE WITH FRIDA}

FRIDA will arrive at the GTC years after similar instruments saw first light on other large telescopes with AO systems. Nevertheless, its combination of high spectral and spatial resolution combined with the light grasp of GTC will be a powerful capability of FRIDA. We would also note that a large amount of observing time is available to the Spanish community (which is the dominant 90\% partner in GTC), which will enable programs that are difficult to carry out with other telescopes.

\section*{ACKNOWLEDGEMENTS}

We are grateful to the Grupo Santander (Spain) for a grant through Encuentros Astrofísicos Blas Cabrera (UNAM-IAC), the GTC project office and our home institutions for their support. Part of this project was carried out under funding of UNAM/PAPIIT grants IT100913 and IT116311, CONACyT grant INFR 2009-01-122664, and CONACyT grant Fronteras 2015-01-226.

\end{document}